\documentclass[%
 reprint,
amsmath,amssymb,
aps,
]{revtex4-2}

\usepackage{graphicx}
\usepackage{dcolumn}
\usepackage{bm}

\begin{document}



\title{Power-law localization in one-dimensional systems with nonlinear disorder under fixed input conditions}

\author{Ba Phi Nguyen}
\email{nguyenbaphi@muce.edu.vn}
\affiliation{Department of Basic Sciences, Mientrung University of Civil Engineering, Tuy Hoa 620000, Vietnam}
\affiliation{Mathematics and Physics Research Group, Mientrung University of Civil Engineering, Tuy Hoa 620000, Vietnam}
\author{Kihong Kim}
\email{khkim@ajou.ac.kr}
\affiliation{Department of Physics, Ajou University, Suwon 16499, Korea}
\affiliation{School of Physics, Korea Institute for Advanced Study, Seoul 02455, Korea}

\begin{abstract}
We conduct a numerical investigation into wave propagation and localization in one-dimensional lattices subject to nonlinear disorder, focusing on cases with fixed input conditions. Utilizing a discrete nonlinear Schr\"odinger equation with Kerr-type nonlinearity and a random coefficient, 
we compute the averages and variances of the transmittance, $T$, and its logarithm, as functions of the system size $L$, while maintaining constant intensity for the incident wave.
In cases of purely nonlinear disorder, we observe power-law localization characterized by 
$\langle T \rangle \propto L^{-\gamma_a}$ and $\langle \ln T \rangle \approx -\gamma_g \ln L$ for sufficiently large $L$.
At low input intensities, a transition from exponential to power-law decay in $\langle T \rangle$ occurs as $L$ increases.
The exponents $\gamma_a$ and $\gamma_g$ are nearly identical, converging to approximately 0.5 as the strength of the nonlinear disorder, $\beta$, increases. Additionally, the variance of 
$T$ decays according to a power law with an exponent close to 1, and the variance of $\ln T$
approaches a small constant as $L$ increases.
These findings are consistent with an underlying log-normal distribution of $T$
and suggest that wave propagation behavior becomes nearly deterministic as the system size increases. 
When both linear and nonlinear disorders are present, 
we observe a transition from power-law to exponential decay in transmittance
with increasing $L$ when the strength of linear disorder, 
$V$, is less than $\beta$. As $V$ increases,
the region exhibiting power-law localization diminishes and eventually disappears when $V$
exceeds $\beta$, leading to standard Anderson localization.
\end{abstract}

\maketitle

\section{Introduction}
\label{sec1} 

Anderson localization is a phenomenon widely observed in linear disordered systems \cite{And,Lee,20,6,23,7}. However, in realistic physical environments such as photonic systems, nonlinear effects significantly alter the transport and localization properties of waves, particularly as their amplitude increases. Despite extensive theoretical and experimental research on Anderson localization
in nonlinear systems \cite{Dev, She, Mol1, kiv, Per, Sch, Lah, Pik, stability, Kop, Fla, Ngu, Vak, Sun, Rica, Ram, Mol2, Sha, Iom, Rias}, the issue has not been satisfactorily resolved. Many questions regarding the interplay between disorder and nonlinearity continue to be open, as highlighted in various reviews \cite{Fis, Lap}.

It is important to distinguish between cases where the intensity of the input wave is fixed while other parameters vary, and those where the output intensity remains constant. The localization behavior exhibits qualitative differences between these two cases, and careful distinction is necessary \cite{Dev,kiv,Ngu,Ram}. While most research on the impact of nonlinearity on Anderson localization has focused on cases with disordered linear potential and nonlinearity as an additional nonrandom effect \cite{Dev, She, Mol1, kiv, Per, Sch, Lah, Pik, stability, Kop, Fla, Ngu, Vak, Sun, Rica}, cases where the disorder also affects the nonlinear terms have rarely been explored \cite{Ram, Mol2, Sha, Iom, Rias}.

In localization studies, which explore wave propagation in disordered media, 
the transmittance $T$ and its logarithm averaged over disorder are crucial quantities,
especially in quasi-one-dimensional cases. The average transmittance indicates the extent of wave penetration through a medium, reflecting the overall effect of disorder on wave transport. The logarithm of transmittance effectively captures the exponential decay of wave transmission due to localization. In conventional exponential localization, the localization length $\xi$ is 
defined by the equation     
$\langle\ln T\rangle=-L/\xi$ in the asymptotic limit,
where $L$ is the thickness of the system
and $\langle\cdots\rangle$ denotes averaging over disorder. These quantities help in quantifying the degree of localization and understanding the underlying physics of wave interaction with disordered structures.

Anderson localization in one-dimensional systems with nonlinear disorder was initially analyzed by Doucot and Rammal, who employed the invariant embedding method to explore theoretical aspects, such as the asymptotic probability distribution \cite{Ram}. Recently, this topic has been revisited by two research groups. One study employed the Helmholtz equation, integrating a Kerr-type nonlinear term with a random coefficient, to numerically study electromagnetic wave propagation in a multilayer structure \cite{Sha}. The results revealed that when disorder is confined solely to the nonlinear term, both wave intensity and transmittance exhibit a power-law decay as the waves penetrate deeper into the medium. Specifically, intensity decays as $1/z$ with increasing penetration depth $z$, while transmittance decreases as $1/L$ with increasing system thickness $L$. Additionally, it was shown that the variances of both intensity and transmittance tend toward zero as $L$ increases. The authors interpreted their findings through the concept of self-induced diffusion. Although not explicitly stated, it is inferred that this work assumed fixed input conditions.

A recent semi-analytical study addressing a similar problem highlighted the differences between cases with fixed input conditions and those with fixed output conditions \cite{Iom}. Under fixed output conditions, the study suggested a power-law localization characterized by a decay in transmittance of $1/L$, whereas fixed input conditions led to conventional exponential decay. The authors proposed that the research discussed in \cite{Sha} was likely conducted under fixed output conditions. Based on these previous studies, we find it crucial to conclusively determine whether systems with purely nonlinear disorder exhibit power-law or exponential localization under fixed input conditions.

In this paper, we present a numerical investigation of wave propagation and localization in one-dimensional lattices under the influence of nonlinear disorder. 
We focus specifically on cases with fixed input conditions, employing a discrete nonlinear Schr\"odinger equation that incorporates Kerr-type nonlinearity with a random coefficient. We compute the averages and variances of the transmittance, $T$, and its logarithm, as functions of the system size $L$, while maintaining a constant intensity, $\vert r_0 \vert^2$, for the incident wave. Our study explores cases with exclusively nonlinear disorder as well as those involving both linear and nonlinear disorders.

In cases of purely nonlinear disorder, we consistently observe a power-law localization characterized by $\langle T \rangle \propto L^{-\gamma_a}$ and $\langle \ln T \rangle \approx -\gamma_g \ln L$ for sufficiently large $L$, contrasting with the exponential localization reported in \cite{Iom}. At sufficiently weak input intensities, we observe a transition from exponential to power-law decay in $\langle T \rangle$ as $L$ increases. 
Surprisingly, the exponents $\gamma_a$ and $\gamma_g$ are almost identical, converging to approximately 0.5 as the strength of the nonlinear disorder increases.
We note that the value of $\gamma_a$ is significantly different from that reported in \cite{Sha}, being approximately half.
Moreover, we demonstrate that the variance of $T$ decays as a power law with an exponent close to 1, which is twice the value of $\gamma_a$, and the variance of $\ln T$ approaches 
a small constant as $L$ increases. We demonstrate that these results are entirely consistent with an underlying log-normal distribution of $T$.
Furthermore, our findings suggest that wave propagation behavior becomes nearly deterministic as the system size increases.

When both linear and nonlinear disorders are present, the localization behavior varies with the relative strengths of each disorder. We observe a shift from power-law to exponential decay in transmittance at a characteristic length when the strength of linear disorder, 
$V$, is less than that of nonlinear disorder, 
$\beta$. As 
$V$ increases, the region of power-law localization diminishes and ultimately vanishes when 
$V$ surpasses 
$\beta$, resulting in standard Anderson localization. 

The remainder of this paper is organized as follows:
In Sec.~\ref{sec2}, we introduce our model and discuss the numerical method.
In Sec.~\ref{sec3}, we detail our numerical results for cases with exclusively nonlinear disorder and for those with both linear and nonlinear disorders.
Finally, in Sec.~\ref{sec4}, we conclude the paper.

\section{Model and method}
\label{sec2} 

As a prototypical model equation, the discrete nonlinear Schr\"odinger equation with a disordered potential finds applications in various areas of physics, such as nonlinear optics and Bose-Einstein condensates \cite{Mol, Kev1, Kev2, Bag}. This equation serves as the basis for exploring phenomena such as Anderson localization in the presence of nonlinearity, turbulence of nonlinear random waves, soliton motion, and more. In one dimension, it takes the form:
\begin{equation}
i\hbar\frac{dC_{n}}{dt}=V_{n}C_{n}+J(C_{n-1}+C_{n+1})+\beta_{n}|C_{n}|^2C_{n},
\label{equation1}
\end{equation}
where $C_{n} (t)$ represents the probability amplitude of finding a particle at the $n$-th site and satisfies the normalization condition $\sum_{n}\vert C_{n}(t)\vert^2=1$. $V_{n}$ denotes the on-site potential at the $n$-th site, and $J$ is the coupling strength between nearest-neighbor sites. $\beta_{n}$ measures the strength of nonlinearity at the $n$-th site. The nonlinear term in the Schr\"odinger equation can result from a mean-field approximation for many-body interactions or from the propagation of waves through nonlinear dielectric media. Henceforth, we will measure all energy scales in units of $J$ and set $J=\hbar=1$, ensuring that energy is dimensionally equivalent to frequency. The stationary solutions of Eq.~(\ref{equation1}) can be expressed in the conventional form: $C_{n}(t)=\psi_{n}e^{-iEt}$, where $E$ represents the energy of an eigenstate.
Substituting this form into Eq.~(\ref{equation1}), we obtain
\begin{eqnarray}
E{\psi_{n}}=V_{n}\psi_{n}+\psi_{n-1}+\psi_{n+1}+ \beta_{n}|\psi_{n}|^2\psi_{n}.
\label{equation2}
\end{eqnarray}
In this study, we will investigate the localization properties of excitations in the presence of either exclusively nonlinear disorder or a combination of both linear and nonlinear disorders. In the former case, only $\beta_n$ is a random variable distributed uniformly over the interval $[-\beta, \beta]$, while in the latter case, both $V_n$ and $\beta_n$ are random variables distributed uniformly over the intervals $[-V, V]$ and $[-\beta, \beta]$, respectively. We note that the disorder average of $\beta_n$ is chosen to be zero, similarly to the models considered in \cite{Sha,Iom}.

In transitioning from Eq.~(\ref{equation1}) to Eq.~(\ref{equation2}), we assumed the existence of a stationary solution for the nonlinear Schr\"odinger equation across all parameters. However, caution is necessary, as previous research has shown that in models where a nonrandom nonlinear term is coupled with a random linear term, the stationary solution can become unstable when the nonlinearity exceeds a certain critical value \cite{stability}. In such cases, these unstable solutions may not be observable in practice. Although theoretically possible, these solutions may not manifest as stable states in experiments due to their transient nature. Understanding the complex dynamics of nonlinear wave propagation remains an important area for further research.

To define the scattering problem in a 1D lattice chain with a finite length $L$, we assume a plane wave is incident from the right side and define the amplitudes of the incident, reflected, and transmitted waves,
denoted by $r_{0}$, $r_{1}$, and $t$, respectively, as follows:
\begin{eqnarray}
\psi_{n}=\left\{\begin{array}{l l}
r_{0}e^{-iq(n-L)}+r_{1}e^{iq(n-L)}, & \quad \mbox{$n\geq L$}\\
te^{-iqn}, & \quad \mbox{$n\leq 0$}
\end{array}\right.,
\label{equation3}
\end{eqnarray}
where the wave number $q$ is related to $E$ by the free-space dispersion relation $E=2\cos q$. In the absence of dissipation, the conservation law $\vert r_{1}\vert^2+\vert t\vert^2=\vert r_{0}\vert^2$ holds, and we choose the overall constant phase for the wave functions so that $t$ is a positive real number.

To numerically calculate the transmittance and reflectance, we first choose an arbitrary positive real number for $t$. Then, using the relationships $\psi_{-1}=te^{iq}$, $\psi_{0}=t$, and $\psi_{1}=E\psi_{0}-\psi_{-1}$, we solve Eq.~(\ref{equation2}) iteratively until we obtain $\psi_{L}$ and $\psi_{L+1}$. Using the relationships $\psi_{L}=r_{0}+r_{1}$ and $\psi_{L+1}=r_{0}e^{-iq}+r_{1}e^{iq}$, we then compute
\begin{eqnarray}
r_{0}=\frac{\psi_{L}e^{iq}-\psi_{L+1}}{e^{iq}-e^{-iq}},~r_{1}=\frac{-\psi_{L}e^{-iq}+\psi_{L+1}}{e^{iq}-e^{-iq}}.
\label{equation4}
\end{eqnarray}
Finally, the transmittance $T$ and the reflectance $R$ are obtained from
\begin{eqnarray}
&&{T}=\left\vert\frac{t}{r_{0}}\right\vert^2=\vert t\vert^2\frac{4\sin^{2}q}{\left\vert\psi_{L}e^{iq}-\psi_{L+1}\right\vert^2},
\label{equation5}\\
&&{R}=\left\vert\frac{r_{1}}{r_{0}}\right\vert^2=\left\vert\frac{\psi_{L}e^{-iq}-\psi_{L+1}}{\psi_{L}e^{iq}-\psi_{L+1}}\right\vert^2.
\label{equation6}
\end{eqnarray}
In contrast to the linear case, where the values of $T$ and $R$ are unaffected by the initial choice for the fixed input $r_{0}$ or output $t$, these choices introduce two distinct problems in the nonlinear case. While the fixed output case is relevant in certain situations, such as when activating a device requires a fixed minimum power at the end of the slab, most practical experiments involve maintaining the strength of the input wave while adjusting other parameters.

A method for calculating wave propagation characteristics in the fixed input case was proposed by us in \cite{Ngu}. Specifically, we first select the values of $L$, $E$ (or $q$), and $\vert r_{0}\vert^2$, along with random configurations of $\beta_{n}$ and $V_{n}$. We then iteratively solve Eq.~(\ref{equation2}) for various initial values of $t$ ($t=0,~\delta,~2\delta,3\delta,\cdots$) until the calculated value of $\vert r_{0}\vert^2$ is sufficiently close to the chosen value. To avoid the bistability or multistability phenomena inherent in the fixed input case, we consider only the first solution to the problem defined by Eqs.~(\ref{equation2}) and (\ref{equation3}). Additionally, it is important to choose the step size $\delta$ appropriately to ensure both the desired accuracy and computational efficiency.

When the nonlinearity is sufficiently strong, it is well-known that bistability or multistability can occur under fixed input conditions, leading to two or more solutions for the given parameters \cite{bs1,bs2,bs3}. In our numerical method described above, a second solution, if it exists, can be obtained by further increasing the initial value of $t$ after the first solution is found, until the calculated value of $\vert r_{0}\vert^2$ closely matches the desired value again. If additional solutions corresponding to multistability exist, this procedure can be repeated until all solutions are identified.
In strongly nonlinear systems, the dependence of $t$ on $\vert r_0\vert^2$ is generally nonmonotonic, meaning that multiple values of $t$ (and consequently different transmittance values) can correspond to a single value of $\vert r_0\vert^2$. Our numerical method can be considered a discretized version of the invariant imbedding method for nonlinear wave propagation, which naturally accounts for the occurrence of multistability. This method is detailed in \cite{bs2}, where the parameters $w$ and $w_0$ shown in Fig.~2(b) correspond to $\vert r_0\vert^2$ and $\vert t\vert^2$ in the present paper. 

\section{Numerical results}
\label{sec3}

We computed disorder-averaged quantities $\langle T\rangle$ and $\langle \ln T\rangle$ as functions of system size $L$, for various values of the incident intensity $\vert r_{0}\vert^2$ and strengths of linear and nonlinear disorders, $V$ and $\beta$, respectively. Although the physical meanings of the parameters $\beta$ and $\vert r_0\vert^2$ are distinct, they are not independent. The results of our model depend on the combined parameter 
$\beta\vert r_0\vert^2$.
All results were obtained by averaging over 500 distinct disorder realizations. The step size for $t$ was set to either $\delta = 10^{-7}$ or $10^{-8}$. The error in the calculated value of $\vert r_{0}\vert^2$ was smaller than $10^{-5}$. Additionally, the excitation wave number was fixed at $q=\pi/2$ (corresponding to the band center $E = 0$) for all results presented in this paper.

\subsection{Nonlinear disorder only}
\label{sec31}

\begin{figure}
\centering
\includegraphics[width=8.6cm]{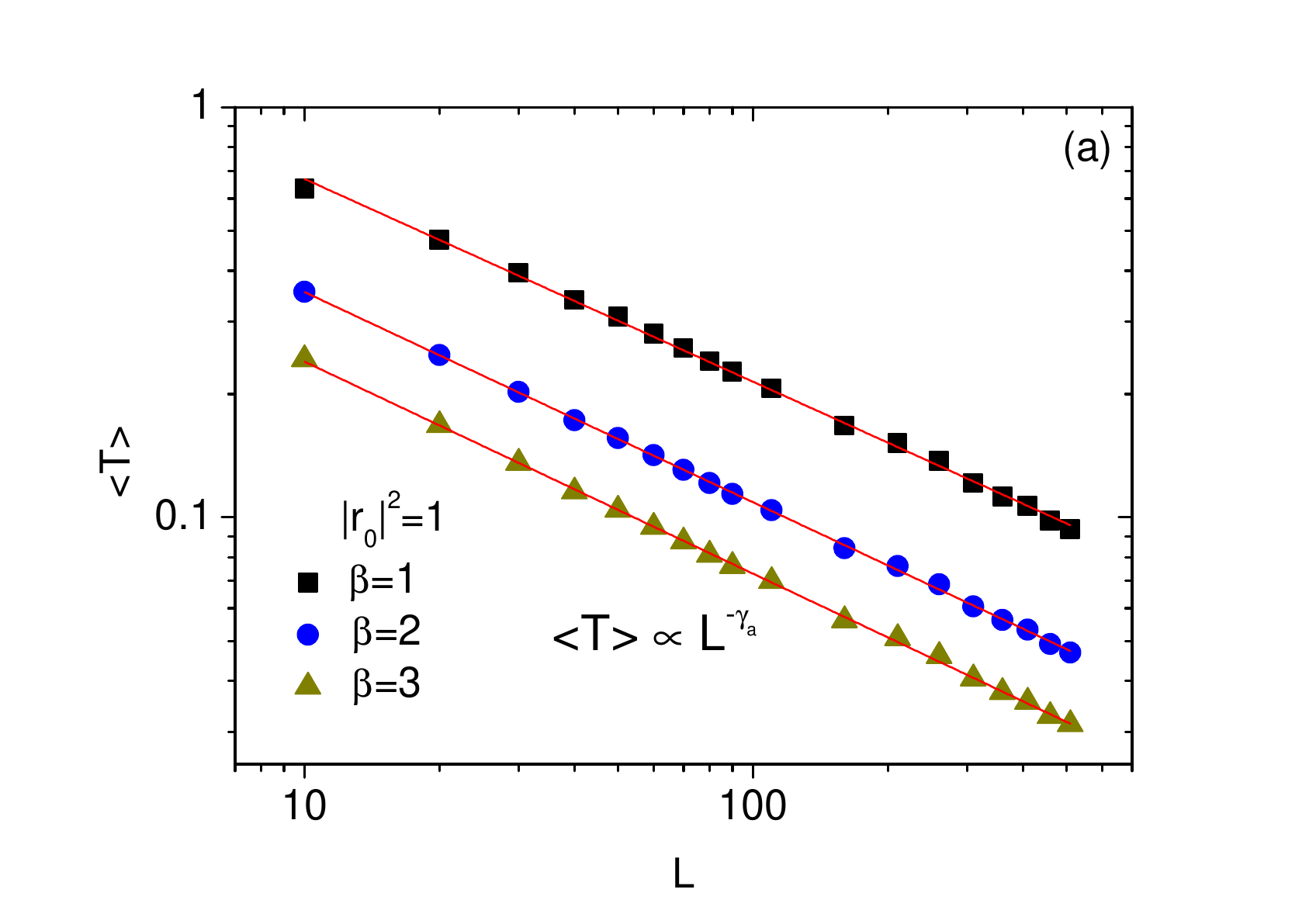}
\includegraphics[width=8.6cm]{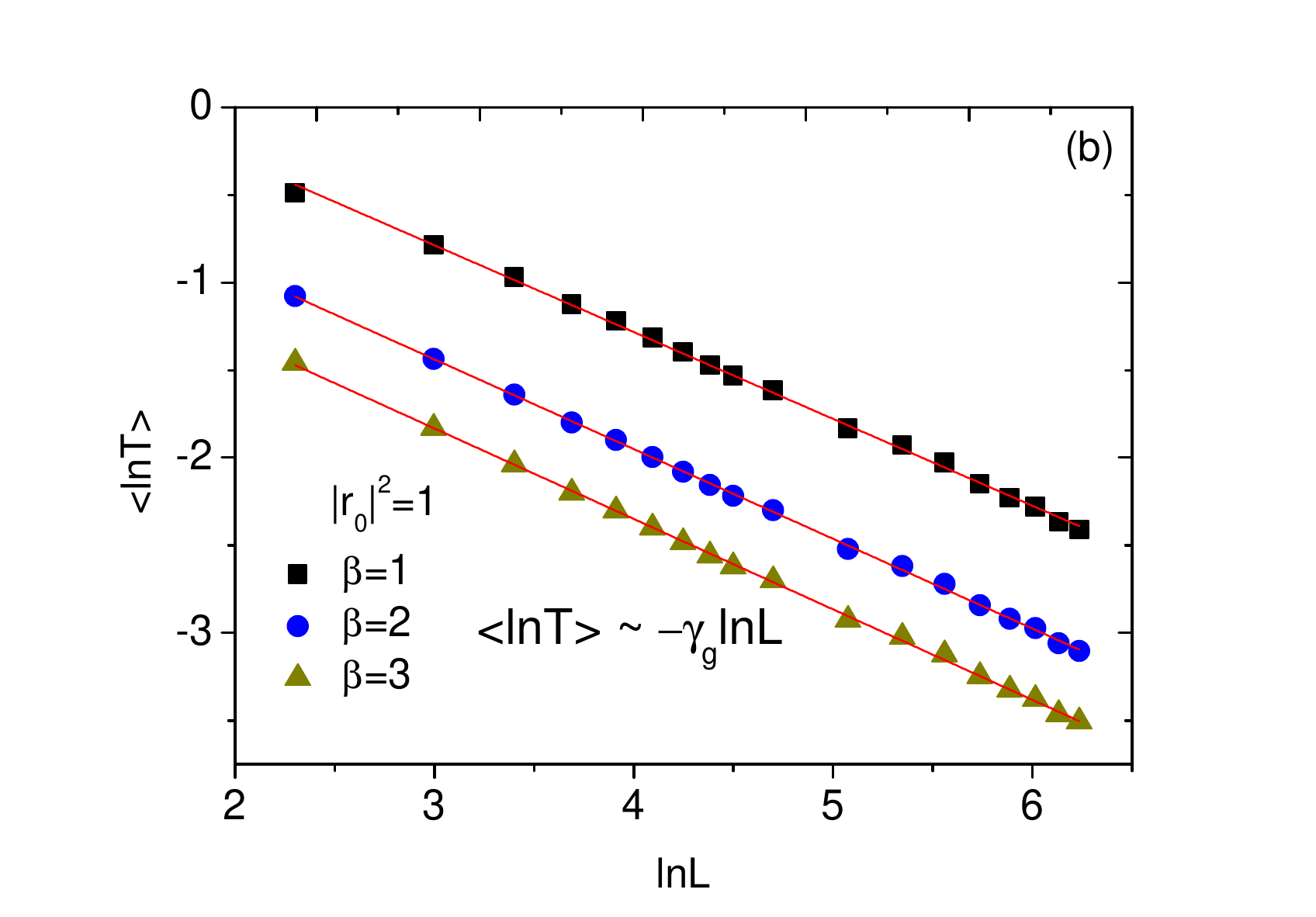}
\caption{
Plots depicting $\langle T\rangle$ and $\langle \ln T\rangle$ as functions of system size $L$ for different nonlinear disorder strengths $\beta$ ($\beta = 1, 2, 3$), with the incident wave intensity fixed at $\vert r_0\vert^2 = 1$. The straight lines in the plots are fits indicative of power-law localization behavior.}
\label{fig1}
\end{figure}

\begin{figure}
\centering
\includegraphics[width=8.6cm]{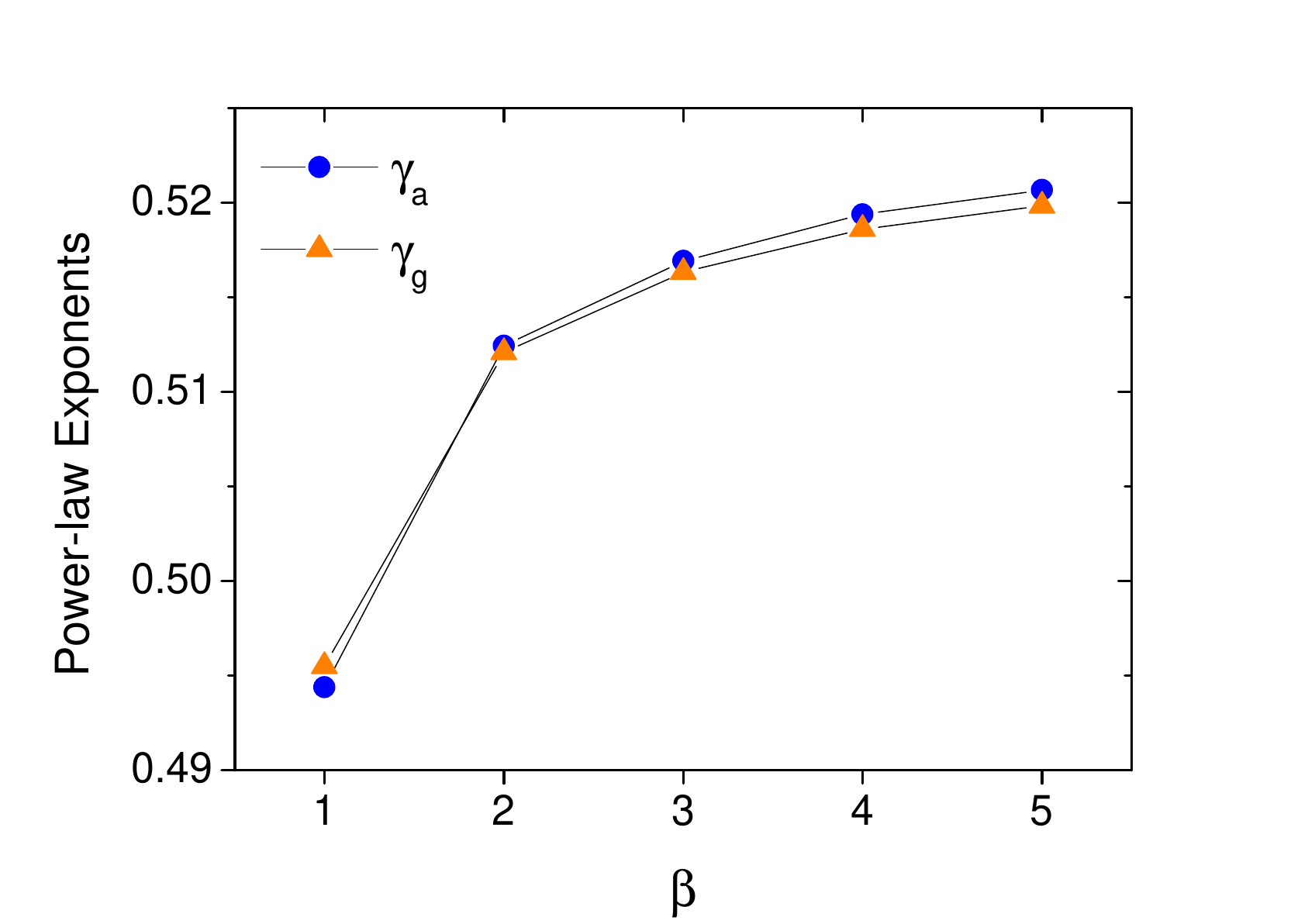}
\caption{Variation of the power-law exponents $\gamma_a$ for $\langle T\rangle$ and $\gamma_g$ for $\langle \ln T\rangle$, plotted as functions of the nonlinear disorder strength $\beta$, with the incident wave intensity $\vert r_0\vert^2$ fixed at 1.}
\label{figs}
\end{figure}

\begin{figure}
\centering
\includegraphics[width=8.6cm]{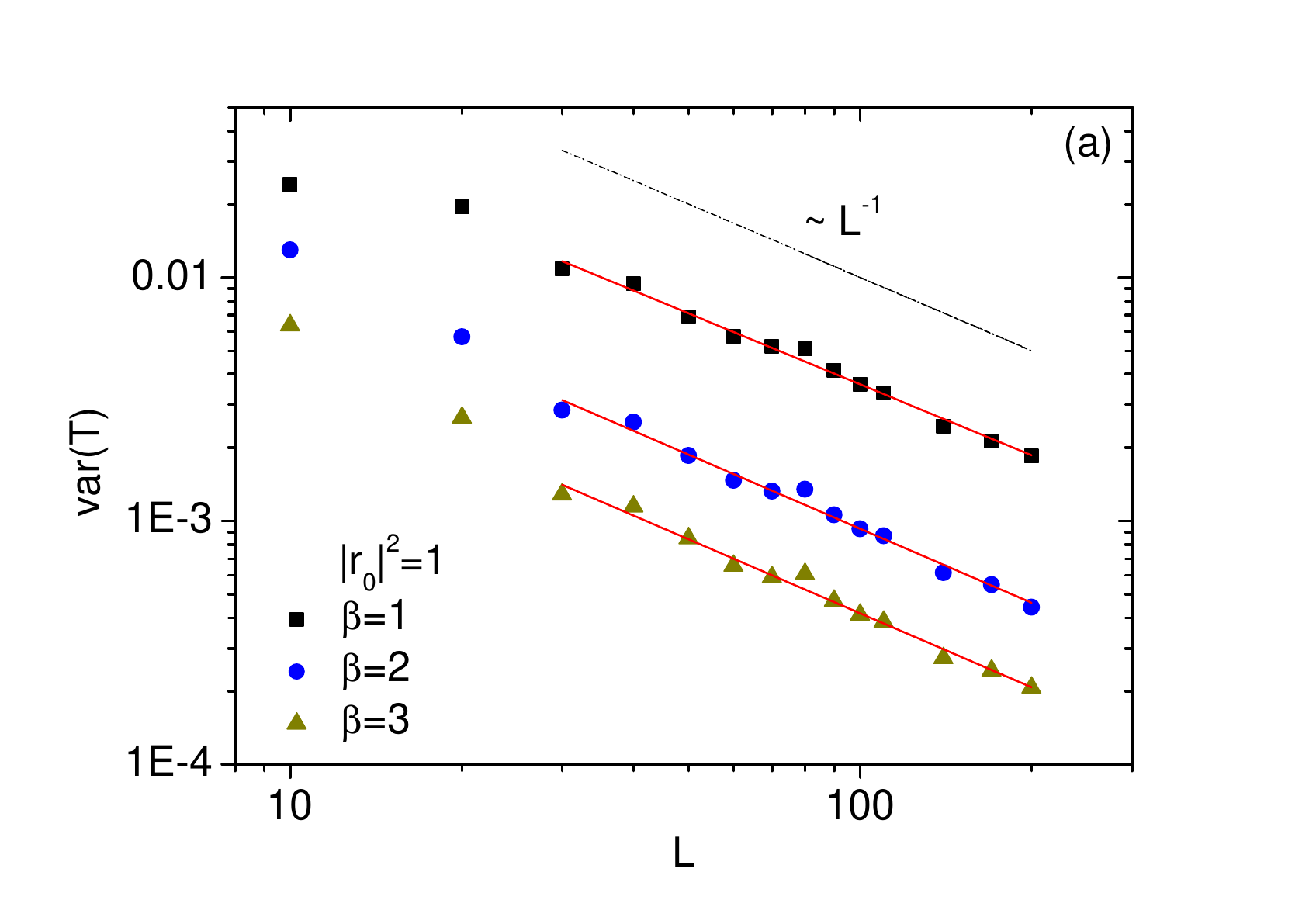}
\includegraphics[width=8.6cm]{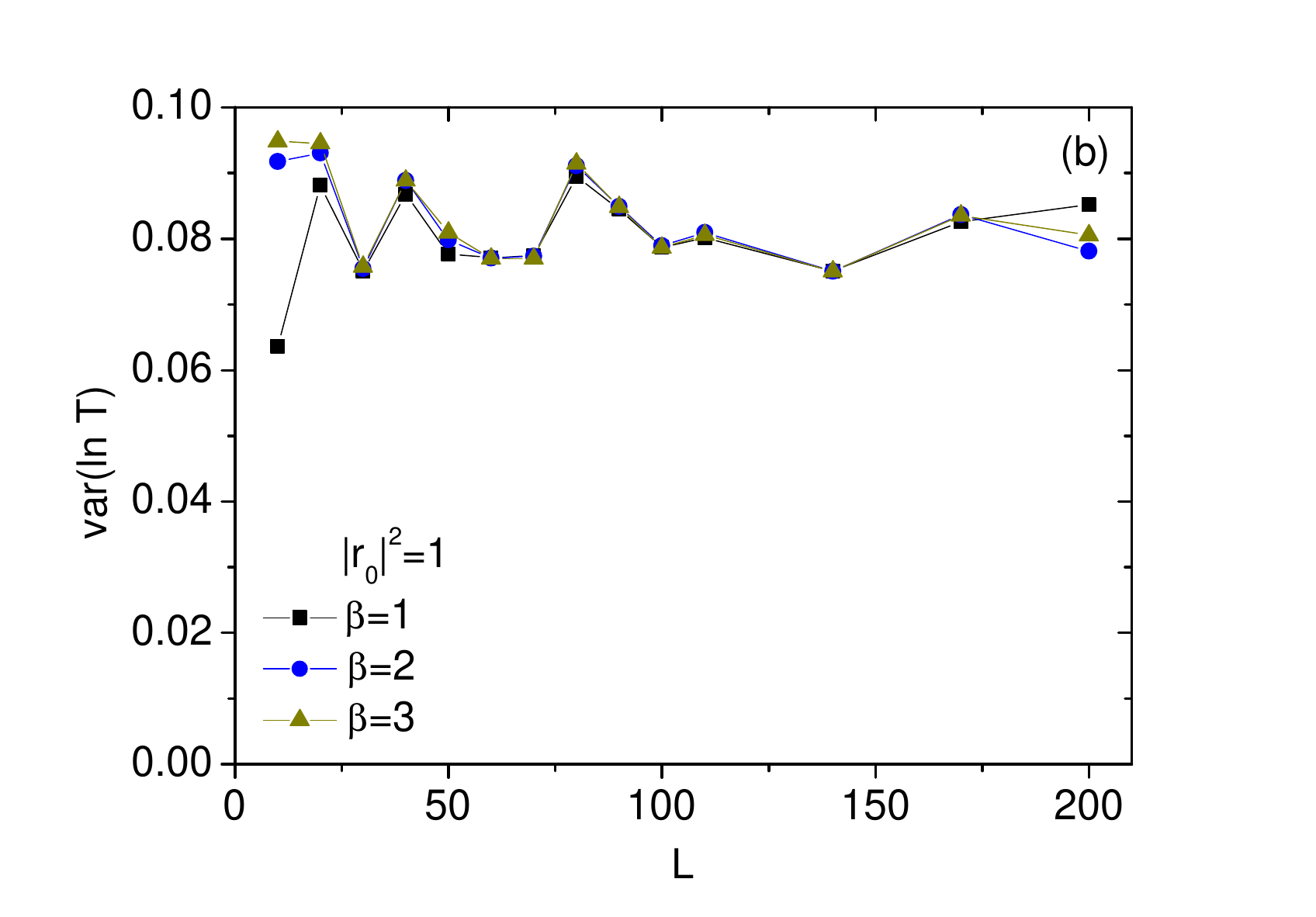}
\caption{(a) Variance of $T$ versus $L$ for $\vert r_{0}\vert^2=1$ and $\beta=1$, 2, 3, shown on a log-log plot. Straight lines from fitting the data in the large $L$ region are compared with the power-law decay of $L^{-1}$. The decay exponents obtained from the fitting are 0.97, 1.01, and 1.01 for $\beta=1$, 2, and 3, respectively. (b) Variance of $\ln T$ versus $L$ for the same parameters as in (a).}
\label{fq}
\end{figure}

We first consider the case where only nonlinear disorder is present. The linear on-site potential $V_n$ is set to zero at all lattice sites, while $\beta_n$ varies randomly within the range $[-\beta, \beta]$ from site to site in a lattice of size $L$. We fix the input intensity $\vert r_{0}\vert^2$ to 1 and calculate $T$ and $\ln T$ using the numerical method described in the previous section. After repeating the calculations for 500 independent random configurations of $\beta_n$, we compute $\langle T \rangle$ and $\langle \ln T \rangle$.

In Fig.~\ref{fig1}, we plot $\langle T\rangle$ and $\langle \ln T\rangle$ versus system size $L$ for $\beta$ values of 1, 2, and 3. For the parameters considered, the decay of transmittance with increasing system size is significantly slower compared to the linear disorder case, which typically shows conventional exponential decay. Here, the decay of transmittance follows a power law, such that $\langle T \rangle \approx AL^{-\gamma_a}$ and $e^{\langle \ln T \rangle} \approx BL^{-\gamma_g}$ in the large $L$ region.
This finding directly contradicts the exponential localization under fixed input conditions reported in \cite{Iom}.

Remarkably, the exponents $\gamma_a$ and $\gamma_g$, derived from curve fitting, are nearly identical, as shown in Fig.~\ref{figs}. These values gradually increase and saturate at approximately 0.52 as the strength of the nonlinear disorder $\beta$ increases. Notably, the exponent $\gamma_a$ reported numerically in \cite{Sha} was 1, which differs significantly from our result. In cases exhibiting power-law decay behavior with a general random distribution of $T$, the exponent $\gamma_g$ typically exceeds $\gamma_a$ by a substantial margin \cite{cdb,bpn1}. The near equivalence of these exponents in our case strongly suggests that, in the asymptotic regime, the fluctuations of 
$T$ are exceedingly small and the underlying probability distribution of 
$T$ follows a log-normal distribution, a point further explored in the Appendix.

The random term $\beta_n\vert \psi_n\vert^2\psi_n$ in Eq.~(\ref{equation2}) becomes significantly smaller than the other terms as the intensity of the wave function decreases. This reduction prevents the accumulation of disorder effects as the propagation length increases. Consequently, in media of sufficient thickness, the influence of the random term becomes negligible well beyond the entry region. As a result, the medium effectively behaves as nonrandom, and wave propagation becomes essentially deterministic. Under these conditions, the fluctuations of $T$ become very small, and
the exponents $\gamma_a$ and $\gamma_g$ become nearly equal.

In the Appendix, we show that when $T$ follows a log-normal distribution and the exponents $\gamma_a$ and $\gamma_g$ are equal, 
the variances of $T$ and $\ln T$ are expressed as
\begin{eqnarray}   
    &&{\rm var}(T)\approx \frac{A^2}{B^2}\left(A^2-B^2\right) L^{-2\gamma_a}\propto L^{-2\gamma_a},\nonumber\\
 &&{\rm var}(\ln T)\approx 2\ln\left(\frac{A}{B}\right)
 \label{eq:var}
\end{eqnarray}
for large $L$.
In Fig.~\ref{fq}, we plot ${\rm var}(T)$ and ${\rm var}(\ln T)$ versus $L$ using the same parameters as in Fig.~\ref{fig1}.
We observe that ${\rm var}(T)$ decays according to a power law with an exponent very close to 1, which is approximately twice the values of 
$\gamma_a$ and $\gamma_g$, in accordance with 
Eq.~(\ref{eq:var}). Additionally, we note that ${\rm var}(\ln T)$ approaches a constant value of about 0.08, 
suggesting that $A/B\approx 1.04$ according to Eq.~(\ref{eq:var}). This result indicates that $\langle T\rangle$ and $e^{\langle \ln T\rangle}$ are nearly identical, and that transmittance behaves almost deterministically as
$L$ becomes large. 

\begin{figure}
\centering
\includegraphics[width=8.6cm]{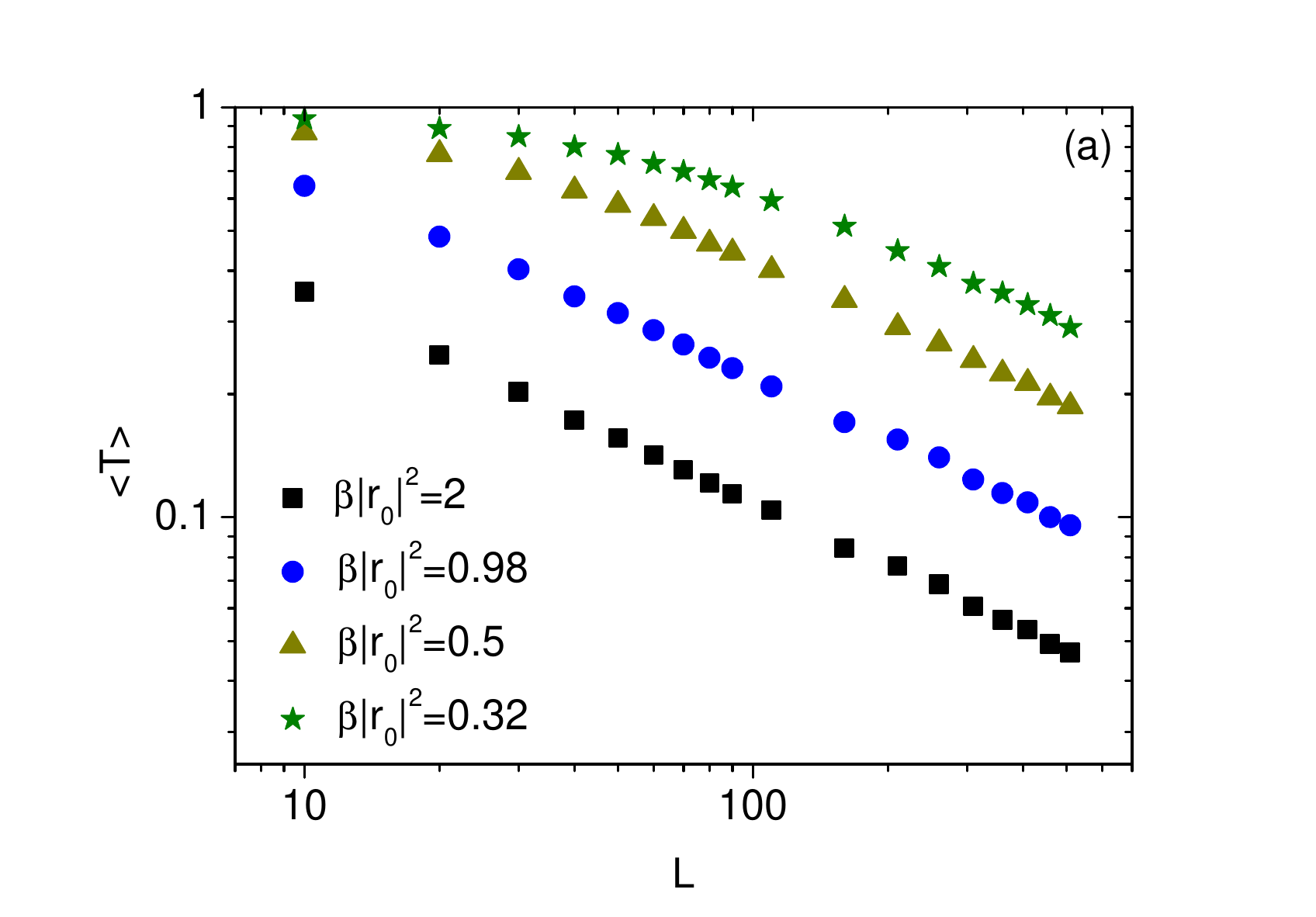}
\includegraphics[width=8.6cm]{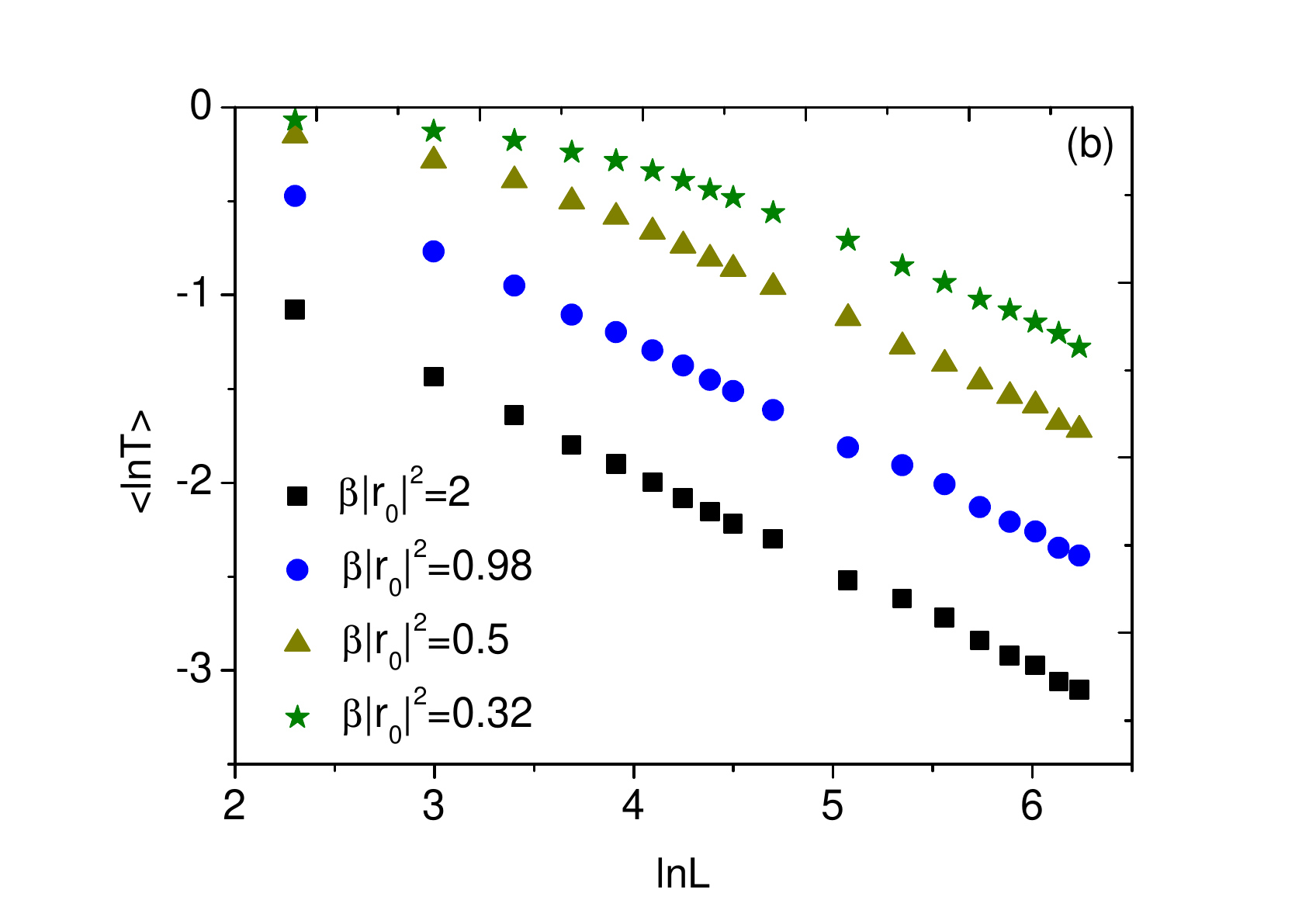}
\caption{Plots showing $\langle T\rangle$ and $\langle \ln T\rangle$ as functions of $L$ for various values of $\beta\vert r_{0}\vert^2$ ($\beta\vert r_{0}\vert^2 = 0.32$, 0.5, 0.98, 2). For a fixed $\beta$, these graphs illustrate the dependence on the incident wave intensity. The plots show two distinct localization behaviors at short and large values of $L$ when the incident intensity is sufficiently low.}
\label{fig2}
\end{figure}

\begin{figure}
\centering
\includegraphics[width=8.6cm]{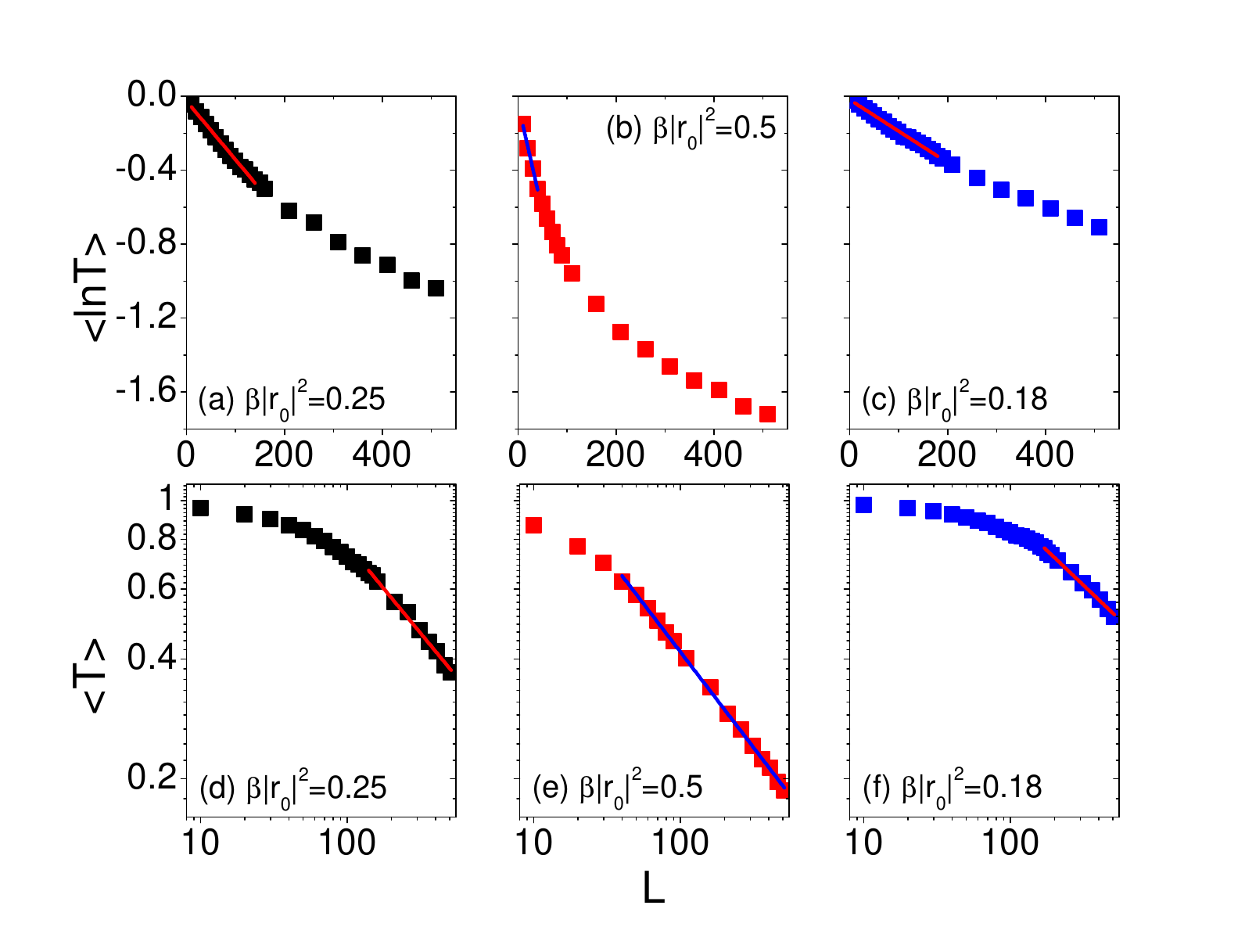}
\caption{Plots illustrating $\langle \ln T\rangle$ and $\langle T\rangle$ versus system size $L$ for various values of $\beta\vert r_{0}\vert^2$. The straight lines represent fitted exponential decay for short $L$ (upper plots) and fitted power-law decay for large $L$ (lower plots). The adjusted R-squared values for these fits exceed 0.994, indicating a high quality fit.}
\label{fig3}
\end{figure}

The observed power-law localization behavior can be qualitatively understood as follows: As excitation waves penetrate deeper into the medium, the intensity of the wave function $\vert \psi_{n}\vert^2$ necessarily decreases. From Eq.~(\ref{equation2}), we find that for a given $\beta$, the effective nonlinear disorder $\beta_{n}\vert\psi_{n}\vert^2$ and, consequently, the backscattering mechanism, weaken as $\vert\psi_{n}\vert^2$ diminishes. This feedback results in a slow, power-law decay of the transmittance. We note that this heuristic argument does not specify the exact functional form of the transmittance as a function of $L$. Remarkably, this power-law decay behavior contrasts sharply with the exponential decay observed in \cite{Mol1,Sch,Lah,Ngu}, where the linear on-site potential is disordered and nonlinearity is introduced by an additional nonrandom term.
In such cases, under fixed input conditions, it has been shown that exponential localization induced by a linear random potential is maintained and enhanced by the presence of uniform nonlinearity \cite{Ngu}.

The power-law localization behavior persists at high incident wave intensities $\vert r_{0}\vert^2$. However, a different behavior emerges as $\vert r_{0}\vert^2$ approaches zero. In Fig.~\ref{fig2}, we present plots of $\langle T\rangle$ and $\langle \ln T\rangle$ versus $L$ for various values of $\beta\vert r_{0}\vert^2$, illustrating the dependence on incident intensity when
$\beta$ is held fixed. We observe that varying the incident intensity distinctly influences localization behavior in the small and large $L$ regions. By selecting appropriate values of $\beta\vert r_{0}\vert^2$, a crossover between exponential and power-law localizations occurs at a specific crossover length. 

To confirm this, in Fig.~\ref{fig3}, we display $\langle T\rangle$ and $\langle \ln T\rangle$ as functions of $L$ for small values of $\beta\vert r_{0}\vert^2$. In all cases examined, conventional Anderson localization is observed in the short $L$ region (upper panels), while power-law localization emerges in the large $L$ region (lower panels). To our knowledge, this transition from exponential to power-law localization has not been previously reported in studies of wave transmission through nonlinear disordered media. This phenomenon can be explained as follows: under conditions of weak incident intensity, the nonlinear disorder effectively mimics linear disorder, resulting in exponential localization in the smaller $L$ region. However, as the wave penetrates deeper into the medium, the decreasing $\vert\psi_{n}\vert^2$ weakens the backscattering mechanism, allowing power-law decay of $\langle T\rangle$ to prevail. The numerical data agree exceptionally well with both exponential and power-law curves, with adjusted R-squared coefficients exceeding 0.994 for all cases studied.

\subsection{Combined linear and nonlinear disorders}
\label{sec32}

\begin{figure}
\centering
\includegraphics[width=8.6cm]{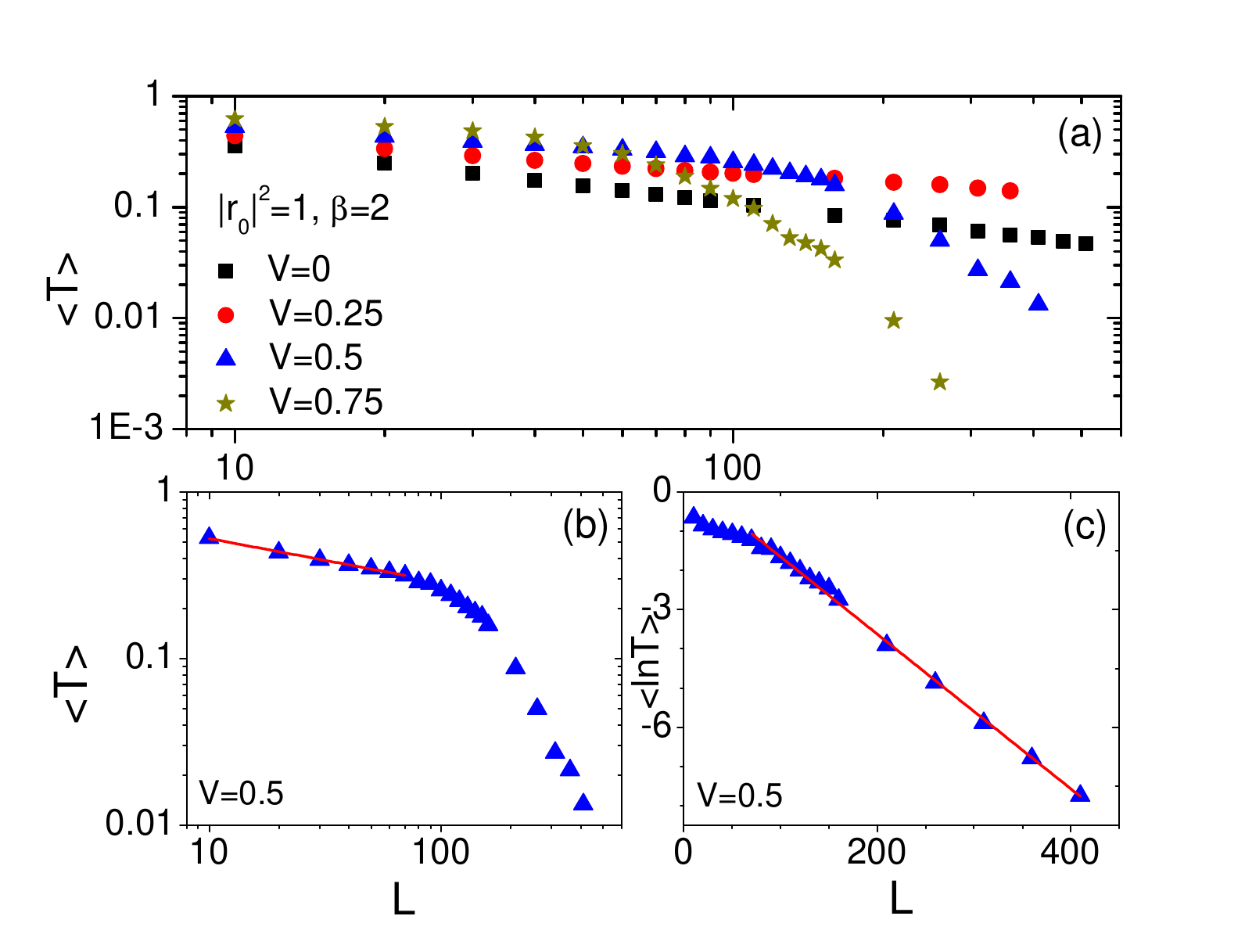}
\caption{(a) Plot of $\langle T \rangle$ versus $L$ for various strengths of linear disorder ($V = 0$, 0.25, 0.5, 0.75), with the incident intensity $\vert r_{0}\vert^2$ and the nonlinear disorder strength $\beta$ fixed at $\vert r_{0}\vert^2$ and $\beta = 2$. Panels (b) and (c) illustrate the transition from power-law to exponential localization, with linear fits to the $\langle T \rangle$ data at small $L$ and $\langle \ln T \rangle$ data at large $L$, respectively, both at $V=0.5$.}
\label{fig4}
\end{figure}
\begin{figure}

\centering
\includegraphics[width=8.6cm]{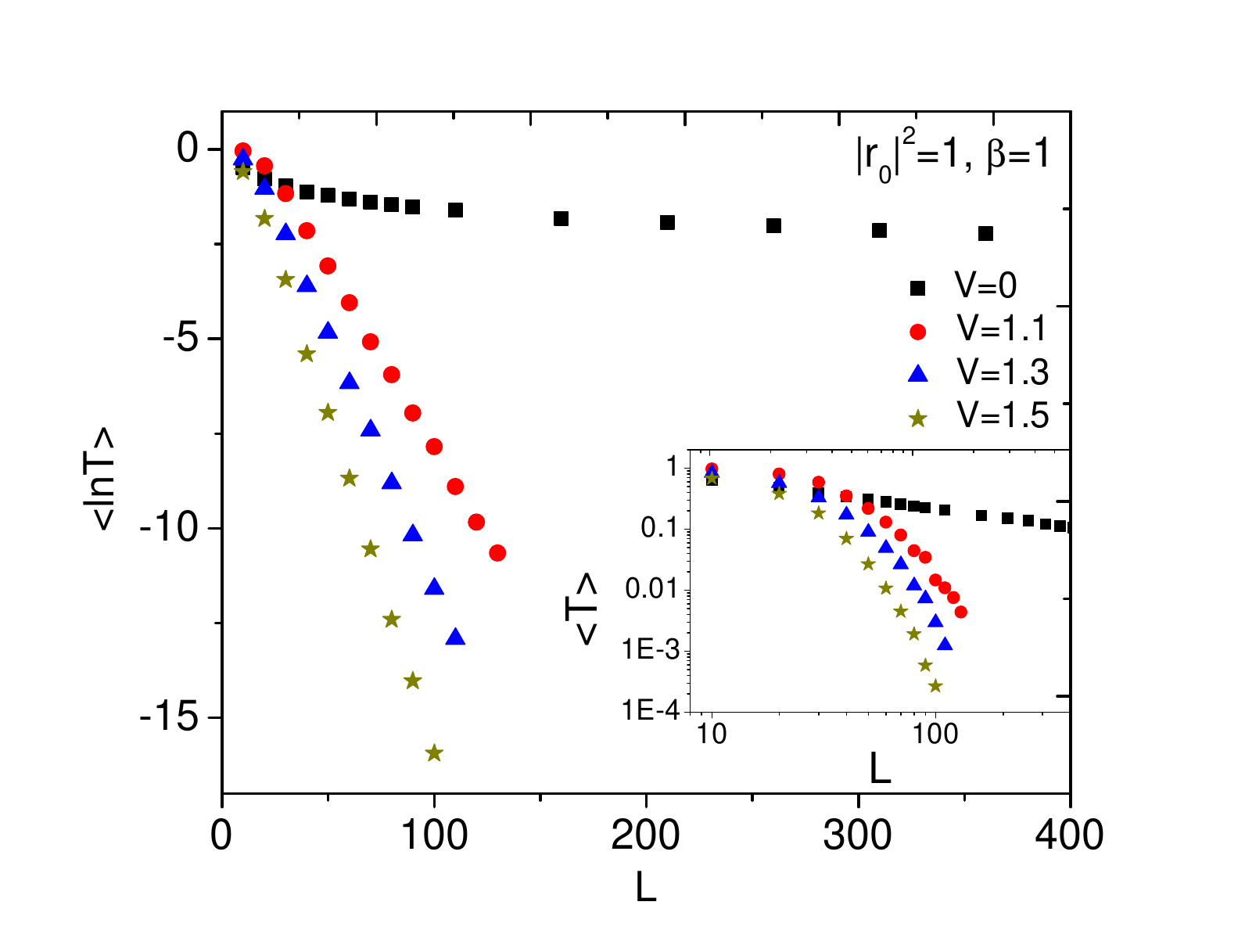}
\caption{Plots of $\langle \ln T \rangle$ and $\langle T \rangle$ versus $L$ for different values of linear disorder strength $V$ ($V=0$, 1.1, 1.3, 1.5), with both the incident intensity $\vert r_{0}\vert^2$ and the nonlinear disorder strength $\beta$ set at 1. Standard exponential Anderson localization is observed for $V > \beta$.}
\label{fig5}
\end{figure}

Achieving purely nonlinear disordered lattices without any linear disorder requires meticulous control over material properties and fabrication processes to ensure that randomness affects only the nonlinear characteristics, while the linear properties remain uniform throughout the system. In practice, such precise isolation of linear and nonlinear properties is challenging, though not impossible. Variations in nonlinear properties often arise due to differences in material composition or structure, which typically also impact the linear properties. For example, in optical media, variations in the nonlinear refractive index due to compositional differences generally correspond with changes in the linear refractive index. Therefore, in this subsection, we consider the combined effects of both linear and nonlinear disorders on wave propagation. These disorders can differently influence wave propagation, potentially leading to distinct behaviors in transmission properties. This subsection focuses on the effects of the relative magnitudes of $\beta$ and $V$.

In Fig.~\ref{fig4}, we display $\langle T\rangle$ and $\langle \ln T\rangle$ as functions of system size $L$ for different values of $V$ ($V=0$, 0.25, 0.5, 0.75), while keeping $\vert r_{0}\vert^2=1$ and $\beta=2$ constant. The power-law localization observed at $V=0$ has been discussed previously. For nonzero $V$ values smaller than $\beta$, nonlinear disorder predominates in the shorter $L$ regions, maintaining power-law localization as shown in Fig.~\ref{fig4}(b). However, as the wave penetrates deeper into the system and its intensity diminishes, the influence of nonlinear disorder decreases, whereas linear disorder remains constant. Consequently, a transition from power-law to exponential decay is observed at a certain characteristic length, which decreases as $V$ increases, as illustrated in Fig.~\ref{fig4}(c).

In Fig.~\ref{fig5}, we present $\langle \ln T\rangle$ and $\langle T\rangle$ as functions of $L$ for $V$ values of 0, 1.1, 1.3, and 1.5, with both $\vert r_{0}\vert^2$ and $\beta$ fixed at 1. Contrary to cases where $V < \beta$ and the transition from power-law to exponential decay occurs gradually with increasing $L$, the transition is more abrupt when $V > \beta$. Here, the influence of nonlinear disorder on wave propagation diminishes rapidly as the wave travels through just a few lattice sites. As linear disorder becomes predominant, the transmittance decays exponentially with increasing system size, leading to pronounced conventional Anderson localization.

\section{Conclusion}
\label{sec4}

In this paper, we have numerically investigated novel localization phenomena in systems with nonlinear disorder. Utilizing a discrete nonlinear Schr\"odinger equation with Kerr-type nonlinearity, we calculated the averages and variances of the transmittance and its logarithm as functions of system size 
$L$, maintaining constant intensity for the incident wave while varying other parameters. In cases of solely nonlinear disorder, we observed a distinct power-law localization phenomenon, where both 
$\langle T\rangle$ and ${\rm var}(T)$ decay following power laws. Further analysis of 
 $\langle \ln T\rangle$ and ${\rm var}(\ln T)$ has led to the intriguing conclusion that the probability distribution of 
$T$ is log-normal, suggesting that wave propagation behavior becomes nearly deterministic as the system size increases. 
These results have been critically compared with previous studies. When both linear and nonlinear disorders are present, the power-law localization behavior attributed to nonlinear disorder competes with the exponential localization due to linear disorder. A transition from power-law to exponential decay in transmittance can occur as 
$L$ increases, particularly if the strength of the nonlinear disorder is sufficiently large. 

The theoretical results presented here can be tested in optical experiments using nonlinear multilayer systems, which are effectively one-dimensional, as discussed in \cite{Sha}. Extending this work to cases involving long-range correlated nonlinear disorder is highly promising, as long-range correlations may induce novel delocalization phenomena.
Another promising research direction is the investigation of the mean square displacement at long times. Similar studies on linear systems exhibiting power-law localization have revealed highly complex dynamic behavior \cite{cdb2}. We expect similarly complex behavior in the present problem, which warrants detailed investigation.

\acknowledgments 
This research was supported through a National Research
Foundation of Korea Grant (NRF-2022R1F1A1074463)
funded by the Korean Government.
It was also supported by the Basic Science Research Program through the National Research Foundation of Korea funded by the Ministry of Education (NRF-2021R1A6A1A10044950).

\appendix
\section{Implications of the log-normal distribution of $T$}
\label{sec_app}

Let us assume that $\ln T$ follows a normal distribution with mean $\mu$ and variance $\sigma^2$. Then, the mean and variance of $T$ can be expressed as
\begin{eqnarray}
\langle T \rangle = e^{\mu + \sigma^2 / 2}, \quad
{\rm var}(T) = e^{2\mu + \sigma^2} \left(e^{\sigma^2} - 1\right).
\end{eqnarray}
Assuming that in the large $L$ region, $\langle T \rangle$ and $\langle \ln T \rangle$ satisfy the equations
\begin{eqnarray}
\langle T \rangle = A L^{-\gamma_a}, \quad e^{\langle \ln T \rangle} = B L^{-\gamma_g},
\end{eqnarray}
we derive the following expressions for $\mu$ and $\sigma^2$:
\begin{eqnarray}
\mu = \ln B - \gamma_g \ln L, \quad \mu + \frac{\sigma^2}{2} = \ln A - \gamma_a \ln L.
\end{eqnarray}
From these, we obtain
\begin{eqnarray}
{\rm var}(\ln T) = \sigma^2 = 2 \ln \left( \frac{A}{B} \right) + 2 (\gamma_g - \gamma_a) \ln L.
\end{eqnarray}
If $\gamma_a$ and $\gamma_g$ are identical, then ${\rm var}(\ln T)$ will approach a constant value of $2 \ln (A / B)$ as $L$ becomes large. Furthermore, the variance of $T$ scales as
\begin{eqnarray}
{\rm var}(T) = \frac{A^2}{B^2}\left(A^2-B^2\right) L^{-2\gamma_a} \propto L^{-2\gamma_a},
\end{eqnarray}
where the power-law decay exponent is twice that of $\langle T \rangle$.

\end{document}